\documentstyle[twoside,natbib,epsf]{article}

\input{ibvs2.sty}

\def\cite#1{\citealt{#1}}
\def\ibvs{Inf. Bull. Var. Stars}
\def\aap{A\&A}
\def\aj{AJ}

\def\apjs{ApJS}

\def\MitVS{MVS}
\def\pasp{PASP}

\begin{document}

\IBVShead{xxxx}{xx April 2002}

\IBVStitletl{IW A\lowercase{nd} is a Z C\lowercase{am}-Type Dwarf Nova}

\IBVSauth{Kato,~Taichi$^1$, Ishioka,~Ryoko$^1$, Uemura,~Makoto$^1$}
\vskip 5mm

\IBVSinst{Dept. of Astronomy, Kyoto University, Kyoto 606-8502, Japan, \\
          e-mail: (tkato,ishioka,uemura)@kusastro.kyoto-u.ac.jp}

\IBVSobj{IW And}
\IBVStyp{UGZ}
\IBVSkey{dwarf nova, classification}

\begintext

   IW And (S~10792) was a blue variable discovered by \citet{mei75iwand}.
\citet{mei80hqandhvandiwand} studied the object spectroscopically
and described that the object seems to be a unique object: in spite of
broad absorption lines of H$\beta$, H$\gamma$ and H$\delta$ resembling
those of an O or early B dwarf or subdwarf, the spectrum was found to be
featureless around H$\alpha$.  \citet{mei80hqandhvandiwand} further
stated that a couple of doubtful emission lines at the limit of detectability
seem to be present.  From 330 observations for the period JD 2440802--46706,
\citet{mei87iwandizandioand} found that the object
spends 72\% of time in an ``inactive" state (15.1--15.3 mag).
The object infrequently showed maximum brightness (18\% of time,
13.7--15.0 mag) and minimum brightness (10\% of time, 15.4--17.3 mag).
\citet{mei87iwandizandioand} stated that such behavior is significantly
different from those of dwarf novae or polars.  More recently,
\citet{liu99CVspec1} obtained a higher quality spectrum, and detected
H$\alpha$ emission line with broad absorption troughs.  Although
\citet{liu99CVspec1} classified the object as a confirmed cataclysmic
variable, the exact nature of the object has not been evident owing to
the lack of dense photometric observations.

   We observed IW And on 55 nights between 2001 December 6 and 2002
March 25.  The observations were done using an unfiltered ST-7E camera
(system close to R$_{\rm c}$) attached to the Meade 25-cm Schmidt-Cassegrain
telescope.  The exposure time was 30 s.  The images were dark-subtracted,
flat-fielded, and analyzed using the Java$^{\rm TM}$-based PSF
photometry package developed by on of the authors (TK).  The differential
magnitudes of the variable were measured against GSC 2811.1573
(Tycho-2 $V$-magnitude 12.05, $B-V$=0.11), whose long-term constancy was
confirmed to 0.10 mag by comparison with GSC 2811.2117 (Tycho-2
$V$-magnitude 11.57, $B-V$=0.69).  The log of observations is summarized
in table 1.

\begin{table}
\begin{center}
Table 1. Nightly averaged magnitudes of IW And \\
\vspace{10pt}
\begin{tabular}{cccccccc}
\hline
Mid-JD$^a$ & Mean mag$^b$ & Error$^c$ & $N$$^d$ &
  Mid-JD$^a$ & Mean mag$^b$ & Error$^c$ & $N$$^d$ \\
\hline
52250.1090 & 2.224 & 0.005 & 31 & 52305.8840 & 2.794 & 0.014 & 31 \\
52250.9993 & 1.973 & 0.006 & 31 & 52306.8847 & 2.721 & 0.017 & 31 \\
52252.1208 & 1.967 & 0.007 & 31 & 52309.9958 & 2.675 & 0.037 & 10 \\
52255.1118 & 2.215 & 0.017 & 31 & 52311.8993 & 2.846 & 0.024 & 31 \\
52255.9875 & 2.325 & 0.043 & 12 & 52312.8972 & 2.786 & 0.030 & 31 \\
52257.0632 & 2.688 & 0.044 & 31 & 52316.9007 & 2.675 & 0.010 & 31 \\
52257.9840 & 3.293 & 0.013 & 31 & 52317.8951 & 2.713 & 0.019 & 31 \\
52259.9757 & 4.398 & 0.048 & 31 & 52318.8986 & 2.639 & 0.017 & 31 \\
52260.9667 & 4.686 & 0.067 & 31 & 52319.8931 & 2.591 & 0.031 & 31 \\
52261.9993 & 4.908 & 0.094 & 31 & 52320.9243 & 2.647 & 0.012 & 31 \\
52262.9660 & 4.415 & 0.034 & 31 & 52323.9382 & 2.887 & 0.179 & 5  \\
52266.9660 & 3.366 & 0.022 & 31 & 52325.9062 & 2.738 & 0.014 & 31 \\
52267.9562 & 4.186 & 0.043 & 31 & 52327.9569 & 2.659 & 0.039 & 31 \\
52270.9486 & 5.017 & 0.119 & 31 & 52329.9188 & 2.866 & 0.013 & 31 \\
52276.9639 & 2.753 & 0.021 & 29 & 52330.9021 & 2.852 & 0.032 & 31 \\
52277.9076 & 2.985 & 0.010 & 31 & 52336.9167 & 2.802 & 0.028 & 31 \\
52279.0493 & 3.347 & 0.159 & 6  & 52337.9062 & 2.732 & 0.044 & 26 \\
52282.9604 & 2.589 & 0.011 & 31 & 52341.9042 & 2.718 & 0.036 & 18 \\
52286.9056 & 2.741 & 0.012 & 31 & 52342.9590 & 2.817 & 0.041 & 12 \\
52291.9104 & 2.852 & 0.042 & 31 & 52344.9076 & 2.875 & 0.046 & 31 \\
52293.8875 & 2.753 & 0.012 & 31 & 52345.9132 & 2.770 & 0.041 & 31 \\
52296.8764 & 2.641 & 0.013 & 31 & 52346.9090 & 2.672 & 0.036 & 31 \\
52297.8875 & 2.704 & 0.027 & 31 & 52348.9111 & 2.800 & 0.087 & 19 \\
52298.8861 & 2.748 & 0.017 & 31 & 52351.9181 & 2.728 & 0.031 & 31 \\
52301.9292 & 2.767 & 0.026 & 31 & 52352.9146 & 2.619 & 0.051 & 31 \\
52302.9847 & 2.769 & 0.038 & 31 & 52353.9153 & 2.829 & 0.079 & 20 \\
52303.8979 & 2.813 & 0.021 & 31 & 52358.9181 & 2.121 & 0.030 & 22 \\
52304.9639 & 2.858 & 0.024 & 31 &            &       &       &    \\
\hline
 \multicolumn{8}{l}{$^a$ JD$-$2400000.} \\
 \multicolumn{8}{l}{$^b$ Relative magnitude to GSC 2811.1573.} \\
 \multicolumn{8}{l}{$^c$ Standard error of nightly average.} \\
 \multicolumn{8}{l}{$^d$ Number of frames.} \\
\end{tabular}
\end{center}
\end{table}

   The resultant light curve is shown in Fig. 1.  The light curve clearly
shows a damping oscllation at the beginning of the observation.  After
that, the object entered a standstill.  The behavior is quite characteristic
to a Z Cam-type dwarf nova entering a standstill (\cite{szk84AAVSO};
\cite{hon98zcam}; \cite{kat01zcam}).  There was even a small hint of
small-amplitude oscillations during the early part of the standstill
which are quite analogous to those of Z Cam (\cite{kat01zcam}) and
HX Peg (\cite{hon98zcam}).  The last observation may indicate that the
object was caught during an outburst from the standstill.
The present observation established that IW And is a previously
unrecognized Z Cam-type dwarf nova.  The ``inactive" state described
in \citet{mei87iwandizandioand} must have been standstills.

\IBVSfig{10cm}{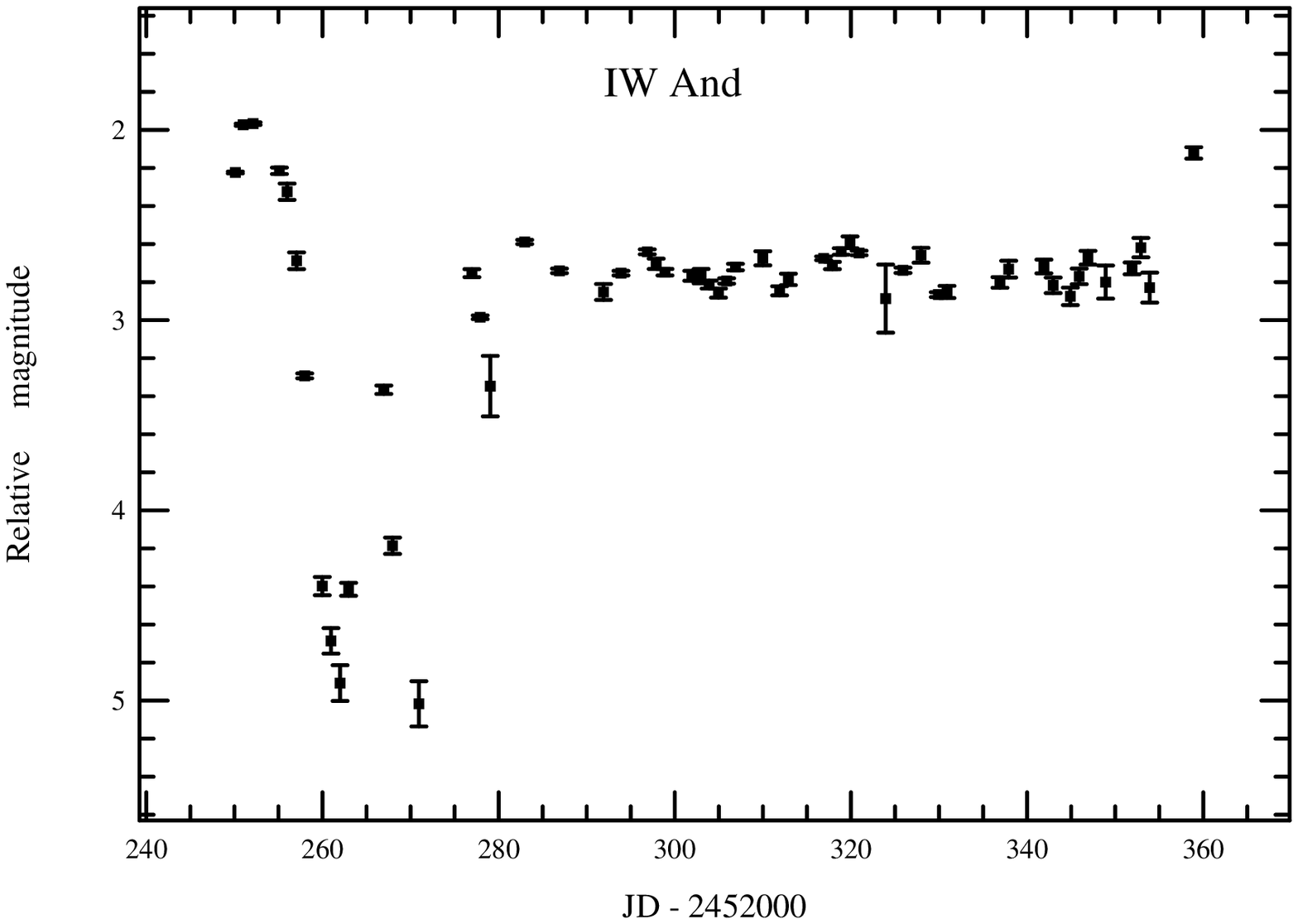}{Light curve of IW And}

   Among Z Cam stars, the duty cycle (nearly 72\%) of standstills is
exceptionally high (cf. the largest duty cycle of well-observed Z Cam stars
is 45\% \cite{opp98zcam}).  Although Z Cam stars have long been understood
as intermediate systems between dwarf novae and novalike (NL) systems
(e.g. \cite{mey83zcam}) in the framework of the disk-instability model
(see \cite{osa96review} for a review), there has been a wide gap between
Z Cam stars and NL systems in terms of the duty cycle of standstills,
which are equivalent to a thermally stable state of NL systems.
IW And is apparently the first object to fill this gap with its large
duty cycle of standstills.  Since such an object is expected to be
provide strong observational constraints to the mechanism of Z Cam
stars (\cite{mey83zcam}, \cite{hon98zcam}, \cite{bua01zcam}), further
continuous observations to precisely determine the pattern of outbursts
and standstills, and spectroscopic observations to determine system
parameters (orbital period, component masses etc.) are strongly encouraged.

\vskip 3mm

This work is partly supported by a grant-in aid (13640239) from the
Japanese Ministry of Education, Culture, Sports, Science and Technology.
Part of this work is supported by a Research Fellowship of the
Japan Society for the Promotion of Science for Young Scientists (MU).

\end{document}